\documentclass[amsmath,amssymb,aps]{revtex4}

\selectfont 

\usepackage{graphicx}       % Include figure files
\usepackage{dcolumn}        % Align table columns on decimal point
\usepackage{bm}             % bold math
\usepackage{psfrag}

\newcommand{\nn}{\nonumber}

\newcommand{\sig}{\sigma}

\begin{document}

\preprint{}

 \title{Scattering of Long-Wavelength Gravitational Waves}

\author{Sam R. Dolan}
 \email{sam.dolan@ucd.ie}
 \affiliation{%
 School of Mathematical Sciences, University College Dublin, Belfield, Dublin 4, Ireland \\
}%

\date{\today}

\begin{abstract}
We consider the scattering of a low-frequency gravitational wave by a massive compact body in vacuum. We apply partial-wave methods to compute amplitudes for the helicity-conserving and helicity-reversing contributions to the cross section, accurate to first order in $M\omega$. Contrary to previous claims, we find that the partial-wave cross section agrees with the cross section derived via perturbation-theory methods.
\end{abstract}

\maketitle

% Paper Plan:
%
% Introduction:
%   

\section{\label{sec:introduction}Introduction}
Gravitational waves -- propagating ripples in spacetime -- are a key prediction of General Relativity (GR). Despite strong indirect evidence for their existence \cite{Hulse-1975,Taylor-1994}, they have yet to be measured directly. This is hardly surprising given the expected amplitude of waves reaching Earth (with a dimensionless strain of $h \sim 10^{-21}$). However, nine decades after the formulation of Einstein's theory, many experimentalists are now optimistic that ``first light'' detections are imminent, at either (existing) ground-based \cite{Waldman-2006} or (future) space-based \cite{Danzmann-2003} interferometers. 

%In recent years, rapid progress has been made in analysing the dynamics of strongly-coupled systems. For example, numerical codes have been developed to simulate the inspiral and merger of a pair of black holes, and their gravitational wave emission. Yet 

In this note, we ask a simple question: when a long-wavelength gravitational wave impinges upon a massive compact body, what is the differential scattering cross section? We will assume that the incident wave is monochromatic, long-lasting, and sufficiently weak that the gravitational field equations may be linearised. Thus, the problem is characterised by a single dimensionless parameter,
\begin{equation}
M \omega = \pi r_S / \lambda ,
\end{equation}
(with units $G = c = 1$) which conveniently expresses the ratio of incident wavelength $\lambda$ to the Schwarzschild horizon $r_S$ of the compact body. In this paper, we concern ourselves only with the long-wavelength regime, in which $M \omega \ll 1$.

It is no surprise to find that this simple question has been asked, and answered, by many authors. The literature on the gravitational scattering of massless waves of various spin ($s=0$, $\tfrac{1}{2}$, $1$, and $2$) is extensive and stretches back over forty years (see \cite{Futterman-1988} and \cite{Frolov-1998} for summaries). Nonetheless, to our knowledge only one paper \cite{Matzner-1977}, written in the late 1970s, tackles this problem via partial-wave methods. The purpose of this note is to revisit and improve Matzner and Ryan's pioneering study \cite{Matzner-1977}.

Over the years, various authors \cite{Westervelt-1971,Peters-1976,Sanchez-1976,Matzner-1977,DeLogi-1977,Doran-2002} have shown that, in the long wavelength limit ($M\omega \ll 1$), the cross section depends on the spin of the scattered field, as follows: 
\begin{align}
\lim_{M \omega \rightarrow 0} \, \left( \frac{1}{M^2} \, \frac{d \sigma}{d \Omega} \right)  &\approx  
\left\{ 
\begin{array}{llll} 
  \frac{1}{\sin^4(\theta / 2)}  & \quad s = 0, & \quad \text{Scalar wave} & \text{[a]} \\ 
  \frac{\cos^2(\theta /2)}{\sin^4(\theta / 2)}  & \quad s = \tfrac{1}{2}, & \quad \text{Neutrino} & \text{[b]}  \\
  \frac{\cos^4(\theta /2)}{\sin^4(\theta / 2)}  & \quad s = 1, & \quad \text{Photon}  & \text{[c]}   \\
  \frac{\cos^8(\theta /2) \, + \, \sin^8(\theta / 2)}{\sin^4(\theta / 2)} & \quad s = 2, & \quad \text{Grav. wave} & \text{[d].} 
\end{array} \right.
\label{csec-low-approx}
\end{align}
It is worth noting that the gravitational result is somewhat anomalous, in that it doesn't follow the same general rule $\left[ d\sig/d\Omega = M^2 \cos^{4s}(\theta/2) / \sin^4(\theta/2) \right]$ as the other fields.

Equation (\ref{csec-low-approx}d) is the conclusion of (at least) three separate studies. The first derivation was carried out by Westervelt \cite{Westervelt-1971}, who applied perturbation theory to the linearised gravitational equations. Peters \cite{Peters-1976} found the same result via a Green's function approach, whilst De Logi and Kov\'acs \cite{DeLogi-1977} applied Feynman diagram techniques. On the other hand, Matzner and Ryan \cite{Matzner-1977} derived a different and more complicated formula by applying partial wave methods. Matzner and Ryan recognised that the lack of agreement between their result and the rest of the literature was surprising, given that, in the case of Coulomb scattering, the lowest-order partial wave cross section and the Born approximation are in exact agreement. 

The primary aim of this work is to show conclusively that Eq. (\ref{csec-low-approx}d) is indeed correct by improving the partial wave analysis of Matzner and Ryan. A secondary aim is to clarify the origin of the `extra' term $\sin^4(\theta/2)$ in the gravitational cross section (\ref{csec-low-approx}d). This term is a direct consequence of the non-conservation of helicity in gravitational-wave scattering. As we shall see, helicity is not conserved because `axial' and `polar' waves are scattered in different ways.

As is well-known \cite{Regge-1957}, first-order perturbations to the Schwarzschild metric may be divided into two classes, according to their behaviour under spatial inversion. \emph{Axial} (or odd) perturbations pick up a factor of $(-1)^{l+1}$ under inversion, whereas \emph{polar} (or even) modes pick up a factor $(-1)^l$. In our discussion, axial/odd modes are referred to as having negative parity ($P=-1$), whereas polar/even modes are said to have positive parity ($P=+1$). As we shall see, partial waves of the same $l$ but opposite parity pick up different scattering phase shifts.

%The implications are discussed in section \ref{sec:discussion}.
\section{Analysis}
The remainder of this note is organised as follows. In section \ref{sec-partial} we briefly recap the results of Matzner \emph{et al.} to write the cross section in terms of amplitudes which are expressed as partial wave series. In \ref{sec-phaseshifts}, we define the phase shifts and discuss their asymptotic values \cite{Poisson-1995} in the long-wavelength regime ($M\omega \ll 1$). In \ref{sec-spin-weighted}, we derive a useful formula for the spin-weighted spherical harmonics. In \ref{subsec-amplitudes}, we show that, in the long-wavelength limit, the partial wave series may be summed to give pleasingly simple results. We conclude with a brief discussion in section \ref{sec:discussion}.

\subsection{\label{sec-partial}Partial Wave Series}
In the late 1970s, Matzner and co-workers \cite{Chrzanowski-1976, Matzner-1977, Matzner-1978, Handler-1980} showed that the differential cross section for the scattering of gravitational waves from a spherically-symmetric compact object can be written as the sum of the square magnitude of two amplitudes,
\begin{equation}
\frac{d \sigma}{d \Omega} = |f(x)|^2 + |g(x)|^2. \label{csec-def}
\end{equation}
These amplitudes may be expressed as partial-wave series,
\begin{align}
f(x) &= \frac{\pi}{i \omega} \sum_{P=\pm 1} \sum_{l=2}^\infty \left[ \exp(2i \delta_{l2}^P) - 1 \right] \,  {}_{-2}Y_l^2(1) \,\, {}_{-2}Y_l^2(x) ,  \label{fg1}  \\
g(x) &= \frac{\pi}{i \omega} \sum_{P=\pm 1} \sum_{l=2}^\infty P (-1)^l \left[ \exp(2i \delta_{l2}^P) - 1 \right] \,  {}_{-2}Y_l^2(1) \,\, {}_{-2}Y_l^2(- x) . \label{fg2}
\end{align}
In these expressions, $\exp(2i \delta_{lm}^\pm)$ are phase factors to be determined from a radial equation, ${}_sY_l^m(x)$ are spin-weighted spherical harmonics, and $x \equiv \cos\theta$, where $\theta$ is the scattering angle. Note the presence of the sum over even and odd parities, $P = \pm 1$. 

The first amplitude $f(x)$ corresponds to (that part of) the interaction which preserves the helicity (i.e. for which the helicity of the scattered wave is the same as the helicity of the incident wave). The second amplitude $g(x)$ corresponds to (that part of) the interaction which reverses the incident helicity. As we see in the next section, the helicity-reversing amplitude is non-zero, because the phase shifts $\delta_{l2}^P$ depend on parity $P$. In this respect, gravitational wave scattering is unlike scalar, neutrino, or electromagnetic scattering.

\subsection{\label{sec-phaseshifts}Phase Shifts}
The phase shifts of odd parity ($P = -1$) may be found from Regge and Wheeler's \cite{Regge-1957} radial equation, 
\begin{equation}
\frac{d^2 R}{d r_\ast^2} + \left[ \omega^2 - V(r) \right] R(r) = 0,  \quad \quad \text{with} \quad V(r) = \left(1 - \frac{2M}{r} \right) \left[ \frac{l(l+1)}{r^2} - \frac{6M}{r^3} \right] , \label{RW-eqn}
\end{equation}
which describes axial perturbations.
Here, $r_\ast$ is a tortoise coordinate defined by $dr / dr_\ast = 1 - 2M/r$.
To find the phase shifts, one must solve this equation subject to the ingoing boundary condition at the horizon [$R(r) \sim \exp(-i\omega r_\ast)$ as $r \rightarrow 2M$]. The asymptotic solution in the far-field is % (\ref{RW-eqn}) may be written as
\begin{equation}
 R(r) \sim A_{\text{in}} \, e^{-i \omega r_\ast} + A_{\text{out}} \, e^{i \omega r_\ast}, \quad \quad \text{as } r \rightarrow \infty ,
\end{equation}
and the phase shifts are determined by the ratio of the ingoing and outgoing coefficients,
\begin{equation}
\exp(2i \delta_{lm}^-) = (-1)^{l+1} A_{\text{out}} /  A_{\text{in}} .
\end{equation}

The phase shifts of even parity $(P=+1)$ may be found by solving Zerilli's radial equation \cite{Zerilli-1970}. It has been shown \cite{Futterman-1988} that the phase shifts of even parity are related to those of odd parity by
\begin{equation}
\exp(2i \delta_{lm}^+) = \frac{(l+2)(l+1)l(l-1) + 12iM\omega}{(l+2)(l+1)l(l-1) - 12iM\omega} \, \exp(2i \delta_{lm}^-). \label{parity-dep}
\end{equation}

Thirty years ago, Matzner and Ryan \cite{Matzner-1977} conducted a partial wave analysis in the low-frequency limit. They assumed the odd-parity phase shift to be approximately
\begin{equation}
\lim_{M\omega \rightarrow 0} \exp(2i \delta_{lm}^-) \approx \frac{\Gamma(l+1-2iM\omega)}{\Gamma(l+1+2iM\omega)}. \label{Matzner-phase}
\end{equation}
More recently, Poisson and Sasaki \cite{Poisson-1995} showed that the exact result for the phase shift in this regime is actually %(ref. , Eq. (3.25))
\begin{equation}
\lim_{M\omega \rightarrow 0} \exp(2 i \delta_{lm}^-) =  e^{-i\Phi} \exp \left( -4iM\omega \, \beta_l \right) ,
\end{equation} 
where
\begin{equation}
\beta_l = \frac{1}{2} \left( \Psi(l+1) + \Psi(l) + \frac{(l-1)(l+3)}{l (l+1)} \right), \quad \quad \text{with} \quad  \Psi(l) \equiv \tfrac{d}{dl}\ln(\Gamma(l)).
\end{equation}
Here, $\Phi \equiv -4M\omega \ln (4M\omega)$ is an overall phase factor which has no effect upon the cross section. The result of Poisson and Sasaki can be rewritten
\begin{align}
\exp(2 i \delta_{lm}^-) &= e^{-i\Phi - 2iM\omega} \exp\left[ -4iM\omega \Psi(l+1) + 8iM\omega / l(l+1) \right] + \mathcal{O}(M^2\omega^2) \\
 &= e^{-i\Phi - 2iM\omega} e^{8iM\omega / l(l+1)}  \frac{\Gamma(l + 1 - 2iM\omega)}{\Gamma(l + 1 + 2iM\omega)}  + \mathcal{O}(M^2\omega^2)  \label{Poiss-Sas-2}
\end{align}
The extra factor of $e^{8iM\omega/l(l+1)}$ present in (\ref{Poiss-Sas-2}) but not in (\ref{Matzner-phase}) proves significant, as we see in section \ref{subsec-amplitudes}. But first, let us briefly digress to study the spin-weighted spherical harmonics.

\subsection{\label{sec-spin-weighted}Spin-Weighted Spherical Harmonics}
To compute the amplitudes (\ref{fg1}) and (\ref{fg2}) we require expressions for the spin-weighted spherical harmonics ${}_{-2}Y_l^2(x)$. These may be found by acting on spherical harmonics of spin-weight zero, ${}_{0}Y_l^0(x) \equiv \sqrt{\frac{2l+1}{4 \pi}} \, P_l(x)$, 
with ladder operators \cite{Goldberg-1967}. The spin-weight is lowered with the operator $\check{\delta}$, and the azimuthal number is raised with $L^+$. These operators are defined by
\begin{align}
\check{\delta} \, {}_sY_l^m(x) = &\left( \sqrt{1-x^2} \, \partial_x - \frac{m + s x}{\sqrt{1-x^2}} \right) {}_sY_l^m(x) = -\sqrt{(l+s)(l-s+1)} \, {}_{s-1}Y_l^m(x), \\
L^+ \, {}_sY_l^m(x) = -&\left(\sqrt{1-x^2} \, \partial_x + \frac{s + mx}{\sqrt{1-x^2}} \right) {}_sY_l^m(x) = \sqrt{(l-m)(l+m+1)} \, {}_sY_l^{m+1}(x).
\end{align}
Here, $\partial_x$ is shorthand for the partial derivative with respect to $x = \cos\theta$. 
By acting with $\check{\delta} L^+ \check{\delta} L^+$ on ${}_{0}Y_l^0(x)$, it is straightforward to show that the spin-weighted harmonics in (\ref{fg1}) and (\ref{fg2}) can be written
\begin{equation}
{}_{-2}Y_l^2 (x) = \sqrt{\frac{2l+1}{4 \pi}} \left( \frac{(1+x)^2 \, \partial_x (1-x) \partial_x \partial_x (1-x) \partial_x P_l(x) }{(l-1)l(l+1)(l+2)} \right) \,   \label{spin-2}.
\end{equation}
Their values in the forward and backward directions are particularly simple,
\begin{equation}
{}_{-2}Y_l^2(1) = \sqrt{\frac{2l + 1}{4 \pi}} \, , \quad \quad {}_{-2}Y_l^2(-1) = 0 .
\end{equation}

%As an aside, we note a useful alternative expression:
%\begin{equation}
%{}_{-2}Y_l^2(x) = \sqrt{\frac{2l+1}{4 \pi}} \left( \frac{6(1-x)^2}{(l-1)l(l+1)(l+2)} \partial_x \partial_x + \frac{4(1-x)}{l(l+1)} \partial_x + 1 \right) P_l(x) .
%\end{equation}

\subsection{\label{subsec-amplitudes}Scattering Amplitudes}
In this section we show that, to first order in $M\omega$, the scattering amplitudes $f(x)$ and $g(x)$ are given by
\begin{align}
M^{-1}\, f(\theta) &= e^{-i\Phi - 2iM\omega} \, \frac{\Gamma(1-2iM\omega)}{\Gamma(1+2iM\omega)} \times \frac{\cos^4(\theta / 2)}{\left[\sin^2(\theta/2)\right]^{1-2iM\omega}}  \; + \mathcal{O}(M\omega) , \label{fresult} \\
M^{-1}\, g(\theta) &= e^{-i\Phi} \times \sin^2(\theta / 2) \; + \mathcal{O}(M\omega) , \label{gresult}
\end{align}
where $e^{-i\Phi}$ is an irrelevant phase factor. 

First, let us consider the helicity-conserving amplitude $f$ defined by (\ref{fg1}). Using results  (\ref{parity-dep}), (\ref{Poiss-Sas-2}), and (\ref{spin-2}), it may be written as
\begin{equation}
f(x) = e^{-i\Phi-2iM\omega} (1+x)^2 \, \partial_x (1-x) \partial_x \partial_x (1-x) \partial_x F(x) , \label{fampl}
\end{equation}
where 
\begin{equation}
F(x) = \frac{1}{2 i \omega} \sum_{l=2}^{\infty} \frac{(2l+1) e^{2 i \Delta_l} P_l(x)}{{(l-1)l(l+1)(l+2)}} ,
\end{equation}
and
\begin{equation}
e^{2 i \Delta_l} \equiv  \left( \frac{(l-1)l(l+1)(l+2)}{(l-1)l(l+1)(l+2) - 12iM\omega} \right) \, \frac{\Gamma(l + 1 - 2iM\omega)}{\Gamma(l + 1 + 2iM\omega)} \, e^{8iM\omega / l(l+1)} + \mathcal{O}(M^2\omega^2) .
\end{equation}
To find the amplitudes in the long-wavelength limit, it is only necessary to keep terms up to first order in $M\omega$. Hence
\begin{align}
\frac{(l-1)l(l+1)(l+2)}{(l-1)l(l+1)(l+2) - 12iM\omega} & \approx \frac{(l-1)l(l+1)(l+2)}{(l-1-2iM\omega)(l+6iM\omega)(l+1-6iM\omega)(l+2+2iM\omega)}, \\
 \text{and} \quad \quad e^{8iM\omega/l(l+1)} & \approx \frac{(l+8iM\omega)(l+1-8iM\omega)}{l(l+1)} .
\end{align}
The phase factor $e^{2 i \Delta_l}$ can then be written
\begin{align}
e^{2 i \Delta_l} & 
  \approx (l-1)l(l+1)(l+2)\, \frac{\Gamma(l - 1 - 2iM\omega)}{\Gamma(l + 3 + 2iM\omega)}.
\label{Delta-phase}
\end{align}
The series $F(x)$ may now be computed with the aid of Eq. (7.127) from Gradshteyn \& Ryzhik \cite{Gradshteyn-1994}, which gives
\begin{align}
\int_{-1}^{1} (1-x)^{\sigma} P_l(x) dx &= \frac{(-1)^l 2^{1 + \sigma} \left[\Gamma(\sigma+1)\right]^2}{\Gamma(\sigma+l+1) \Gamma(1 + \sigma - l)} , \quad \quad \text{Re}(\sigma) > -1 , \nn \\
 &= 2^{1 + \sigma} \sigma (\sigma - 1) \frac{\Gamma(1+\sigma)}{\Gamma(2-\sigma)} \frac{\Gamma(l-\sigma)}{\Gamma(l+1+\sigma)} .
\end{align}
To find $F(x)$ we substitute $\sigma = 1 + 2iM\omega$ into the above formula and compare with $\int_{-1}^{1} F(x) P_l(x) dx$. This implies that
\begin{equation}
F(x) = M \, \frac{2^{-(1 + 2iM\omega)} }{(2iM\omega)^2 (1+2iM\omega)^2} \frac{\Gamma(1 - 2iM\omega)}{\Gamma(1 + 2iM\omega) } (1 - x)^{1 + 2iM\omega}  .
\label{Fx}
\end{equation}
Plugging (\ref{Fx}) into (\ref{fampl}) and taking four derivatives yields the pleasingly simple result
\begin{equation}
f(x) = M e^{-i\Phi - 2iM\omega} \frac{\Gamma(1-2iM\omega)}{\Gamma(1+2iM\omega)} \, \frac{\left[\tfrac{1}{2}(1+x)\right]^2}{\left[\tfrac{1}{2}(1-x)\right]^{1-2iM\omega}}
\end{equation}
which is the same as (\ref{fresult}). 

The helicity-reversing amplitude $g(x)$ defined in (\ref{fg2}) may be written
\begin{equation}
g(x) = \frac{e^{-i\Phi} \sqrt{4\pi}}{2i\omega} ( 12iM\omega ) \sum_l \frac{ \sqrt{2l+1} \, (-1)^l}{(l-1)l(l+1)(l+2)} \, e^{-4iM\omega \beta_l} \, {}_{-2}Y_l^2(-x) \,
\label{gx2}  
\end{equation} 
The denominator of this expression ensures that the series converges quickly. We are only interested in the amplitude to first order in $M\omega$, and there is already a factor of $12iM\omega$ in (\ref{gx2}). Hence it is justified to take $e^{-4iM\omega \beta_l} \sim 1 + \mathcal{O}(M\omega)$.

In order to compute $g(x)$, let us first consider the integral
\begin{align}
I &= \int_{-1}^{1} (1-x) \, {}_{-2}Y_l^2(-x) dx \nn \\
  &= \sqrt{\frac{2l + 1}{4\pi}} \, \frac{(-1)^l}{ (l-1)l(l+1)(l+2)} \int_{-1}^1 (1-x)^3 \partial_x (1+x) \partial_x \partial_x (1+x)\partial_x P_l(x) .
\end{align}
Integrating by parts four times we obtain
\begin{align}
I &= \sqrt{\frac{2l+1}{4\pi}} \, \frac{(-1)^l}{(l-1)l(l+1)(l+2)}    \left[ 6(3x-1)(1+x)P_l(x) \right]_{-1}^1 \nn   \\
  &= 24 \, \sqrt{\frac{2l+1}{4\pi}} \, \frac{(-1)^l}{(l-1)l(l+1)(l+2)}
\label{Iint}
\end{align}
for $l \ge 2$. 
Applying the orthogonality relation for the spin-weighted spherical harmonics,
\begin{equation}
\int_{-1}^{1} {}_{-2}Y_l^2(x) \, {}_{-2}Y_{l^\prime}^2(x) dx = \frac{1}{2\pi} \, \delta_{ll^\prime} ,
\end{equation}
we conclude that 
\begin{align}
\int_{-1}^1 g(x) \, {}_{-2}Y^2_l(-x) dx = 12 M e^{-i\Phi} \sqrt{\frac{2l+1}{4\pi}} \frac{(-1)^l}{(l-1)l(l+1)(l+2)}.
\end{align}
Hence, by comparison with (\ref{Iint}), 
\begin{equation}
g(x) = \tfrac{1}{2} M e^{-i\Phi} (1-x),
\end{equation}
which is the same as result (\ref{gresult}).

\section{\label{sec:discussion}Discussion and Conclusion}
In the preceding sections we have computed the partial-wave scattering amplitudes defined by Matzner and co-workers \cite{Matzner-1977, Matzner-1978} in the long-wavelength limit ($M\omega \ll 1$). Substituting (\ref{fresult}) and (\ref{gresult}) into (\ref{csec-def}) we conclude that
\begin{equation}
\frac{d\sig}{d\Omega} = M^2 \, \frac{ \sin^8 (\theta/2) + \cos^8(\theta/2) }{\sin^4(\theta / 2)}. \label{final-result}
\end{equation}
The partial-wave result is therefore consistent with all previous studies \cite{Westervelt-1971, Peters-1976, DeLogi-1977}, which reach the same result by applying perturbation-theory methods.

It is worth remarking that, unlike other polarised waves (i.e. neutrino and photon waves), the scattering cross section of the gravitational wave (\ref{final-result}) is non-zero in the backward direction ($\theta = \pi$). As we have seen, this is because the phase shifts in the partial wave series are parity-dependent (Eq. \ref{parity-dep}). This implies the existence of a helicity-reversing amplitude $g(x)$. As noted by De Logi and Kov\'acs \cite{DeLogi-1977}, ``if the incident radiation is in a pure helicity state, the backscattered ($\theta=\pi$) radiation must have the opposite helicity''. %The gravitational field linearised on a non-flat background does \emph{not} have the usual geometric properties of a pure spin-two field. 

Finally, we note that a spherically-symmetric interaction will not induce a net polarisation in an initially unpolarised beam. Right- and left-circular polarisations are scattered in the same way (Eq. \ref{final-result}). If the scattering body is rotating, the spherical symmetry is broken. It has been suggested \cite{DeLogi-1977} that, to first order in $M\omega$, polarisation is \emph{not} induced by a rotating scatterer. However, outside this regime (i.e. when $M\omega \sim 1$) we would definitely expect to see some polarisation effects caused by rotation. For instance, if the incident wave were to impinge along the rotation axis of a Kerr black hole, then the co-rotating polarisation would be enhanced by the superradiance effect \cite{Handler-1980}. We hope to make a numerical invesigation of scattering from a rotating black hole in the near future.

\begin{acknowledgments}
I would like to thank Marc Casals and Barry Wardell for proof-reading the manuscript and for helpful suggestions. Financial support from the Irish Research Council for Science, Engineering and Technology (IRCSET) is gratefully acknowledged.
\end{acknowledgments}

%Numerous studies of black hole scattering have been made over the last forty years. The most relevant for our purposes Pioneering contributions include the works of Hildreth \cite{Hildreth-1964}, Matzner and co-workers \cite{Matzner-1968, Matzner-1977, Matzner-1978, Handler-1980} and Sanchez \cite{Sanchez-1976, Sanchez-1977, Sanchez-1978a, Sanchez-1978b}. Early studies focus on the simplest setup: scalar (Klein-Gordon) waves impinging on a non-rotating (Schwarzschild) hole. Later works extend the analysis to fields of higher spin: that is, fermionic \cite{Dolan-2006}, electromagnetic \cite{Fabbri-1975} and gravitational \cite{Futterman-1981} fields. Many authors found regular oscillations in the resulting cross-section-versus-scattering-angle plots. The physical origin of the oscillations was clarified by DeWitt-Morette \cite{DeWitt-Morette-1984}, Zhang \cite{Zhang-1984} and others \cite{Anninos-1992}. These authors showed that the interference of null rays passing close to the unstable photon orbit at $r = 3M$ resulted in a regular diffraction pattern. Thus the study of large-angle scattering cross sections could reveal details of the internal structure of the hole.

\bibliographystyle{apsrev}

%\bibliography{year1}

\end{document}